\documentclass[twocolumn,showpacs,prl,preprintnumbers,superscriptaddress,amsmath,amssymb]{revtex4-1}

\usepackage{txfonts}
\usepackage{graphicx}
\usepackage{epstopdf}
\usepackage{float}
\usepackage{overpic}
\usepackage{dcolumn}
\usepackage{bm}
\usepackage{gensymb}
\usepackage{textcomp}
\usepackage{chemformula}
\usepackage{mhchem}
\usepackage[%
	breaklinks=true,
	colorlinks=true,
	linkcolor=blue,
	urlcolor=blue,
	citecolor=blue,
	]{hyperref}
\usepackage{color}
\usepackage{mathtools}
\usepackage{lineno}
\begin{document}

\title{Decoding Molecular Geometries in Coulomb Explosion Imaging via Physics-Informed Deep Neural Network}

\author{Xingyu Guo}
\affiliation{Hefei National Research Center for Physical Sciences at the Microscale and Department of Modern Physics, University of Science and Technology of China, Hefei, 230026, China}

\author{Enliang Wang}
\email{elwang@ustc.edu.cn}
\affiliation{Hefei National Research Center for Physical Sciences at the Microscale and Department of Modern Physics, University of Science and Technology of China, Hefei, 230026, China}

\author{Wenguang Wu}
\affiliation{Hefei National Research Center for Physical Sciences at the Microscale and Department of Modern Physics, University of Science and Technology of China, Hefei, 230026, China}

\author{Zhaopeng Xing}
\affiliation{Hefei National Research Center for Physical Sciences at the Microscale and Department of Modern Physics, University of Science and Technology of China, Hefei, 230026, China}

\author{Tuo Liu}
\affiliation{Hefei National Research Center for Physical Sciences at the Microscale and Department of Modern Physics, University of Science and Technology of China, Hefei, 230026, China}

\author{Chunkai Xu}
\affiliation{Hefei National Research Center for Physical Sciences at the Microscale and Department of Modern Physics, University of Science and Technology of China, Hefei, 230026, China}
\affiliation{Hefei National Laboratory, Hefei, Anhui, 230088, China}

\author{Xu Shan}
\affiliation{Hefei National Research Center for Physical Sciences at the Microscale and Department of Modern Physics, University of Science and Technology of China, Hefei, 230026, China}

\author{Artem Rudenko}
\affiliation{J.R. Macdonald Laboratory, Department of Physics, Kansas State University, Manhattan, KS, USA.}
\author{Daniel Rolles}
\affiliation{J.R. Macdonald Laboratory, Department of Physics, Kansas State University, Manhattan, KS, USA.}

\author{Jing Chen}
\affiliation{Hefei National Research Center for Physical Sciences at the Microscale and Department of Modern Physics, University of Science and Technology of China, Hefei, 230026, China}
\affiliation{Hefei National Laboratory, Hefei, Anhui, 230088, China}

\author{Xiangjun Chen}
\email{xjun@ustc.edu.cn}
\affiliation{Hefei National Research Center for Physical Sciences at the Microscale and Department of Modern Physics, University of Science and Technology of China, Hefei, 230026, China}
\affiliation{Hefei National Laboratory, Hefei, Anhui, 230088, China}

\date{\today}

\begin{abstract}
Determining the absolute configuration of gas-phase molecules in position-space has long been a fundamental challenge in molecular physics. While strong-field-induced Coulomb explosion imaging (CEI) has emerged as a powerful tool for probing molecular stereochemistry in momentum-space, reconstructing the original three-dimensional structure of polyatomic molecules remains a long-standing challenge due to the inherent complexity of multidimensional inversion. Here, we introduce a deep learning framework that bridges this gap by directly recovering position-space molecular structures from Coulomb explosion momentum patterns. Our approach combines CEI simulations with a neural network trained to establish the mapping between momentum-space Newton plots and real-space geometries. The trained model demonstrates high fidelity in reconstructing the structure of  CHF$_3$ from experimental CEI data. This generalizable framework 
can not only be extended to other molecular systems but also opens avenues for time-resolved structural analysis of molecular dynamics.
\end{abstract}

\maketitle
Determining molecular structures in the gas phase represents a fundamental challenge in modern physics. While X-ray diffraction and nuclear magnetic resonance spectroscopy have revolutionized structural analysis in condensed phases, their application to isolated molecules remains severely limited by low target density. Only recently, the high-intensity X-ray free-electron laser  has allowed a diffraction experiment on a gas-phase target. Alternatively, Coulomb explosion imaging (CEI) has emerged as a powerful tool, providing direct access to molecular geometry in momentum space through the measurement of fragment ion momenta following rapid multielectron ionization \cite{Kanter_PRA1979, vager_Science1989, Herwig_Science2013}. The CEI technique was first demonstrated by Kanter et al. in 1979 \cite{Kanter_PRA1979} through foil-induced dissociation of diatomic molecular ions, e.g. H$_2^+$ and HeH$^+$.  The methodology was subsequently extended to polyatomic systems by Vager et al. \cite{vager_Science1989}, who resolved the stereochemistry of hydrocarbon molecules, and later advanced by Herwig et al.  \cite{Herwig_Science2013} and Pitzer et al. \cite{Pitzer2013Science} to achieve absolute configuration of a chiral molecule. In addition to foil-induced ionization, CEI has been established by various ionization methods, including highly charged ion impact \cite{Neumann2010PRL, Yuan2024PRL}, electron impact \cite{Wang2015PRA, Wang2021PRL}, x-ray/free-electron lasers \cite{Pathak2020JCPL, Boll2022NP, Li2022PRR, Richard_Science2025, Green_JACS2025}, as well as tabletop strong-field femtosecond lasers \cite{Pitzer2013Science, kunitski2015science, Cheng2023PRL, Lam2024PRL, Wang2023Arxiv}.

These pioneering works established the foundation for momentum-based structure determination which correlates fragment ion trajectories with initial molecular geometries. The core principle relies on rapid electron stripping, which projects the initial molecular geometry to momentum space through Coulomb repulsion in the resulting highly charged ion, with the momentum vectors of the fragment ions encoding critical information about the original molecular configuration \cite{Schouder2022ARPC}. 
While back-projection methods have successfully reconstructed bond lengths in simple diatomic molecules \cite{Schmidt2012PRL, Zeller2016PNAS}, position-space reconstruction of polyatomic systems remains an outstanding challenge due to two fundamental limitations: first, the multidimensional complexity of the inversion, as the explosion dynamics of polyatomic systems involve coupled many-body interactions that cannot be adequately addressed by simplistic back-projection algorithms; and second, the inherently ill-posed nature of the reconstruction process, where distinct initial geometries may produce nearly identical momentum distributions. A notable exception is the three-body reconstruction of helium trimer Efimov states, achieved by comparing Dalitz plots in both momentum and position spaces to map the unique geometric configuration of this exotic quantum system \cite{kunitski2015science}, though such sophisticated methods still struggle with filtering out non-bijective correspondence events and scaling to more complex molecular systems.
Recent advances in deep learning, however, offer a transformative solution, as deep neural networks have demonstrated remarkable success in solving intricate scientific problems by establishing direct mappings between observable quantities and underlying physical parameters \cite{Zhang2018PRL,westermayr2022NC, Tenachi2023APJ}. These developments suggest that deep learning approaches could provide a critical pathway to overcoming the limitations of traditional CEI reconstruction, enabling robust position-space recovery through nonlinear pattern recognition, bypassing the explicit modeling of many-body quantum dynamics, and filtering degenerate solutions via training with physical constraints.

\begin{figure}[!t]
	\centering
	\includegraphics[width=7.5cm]{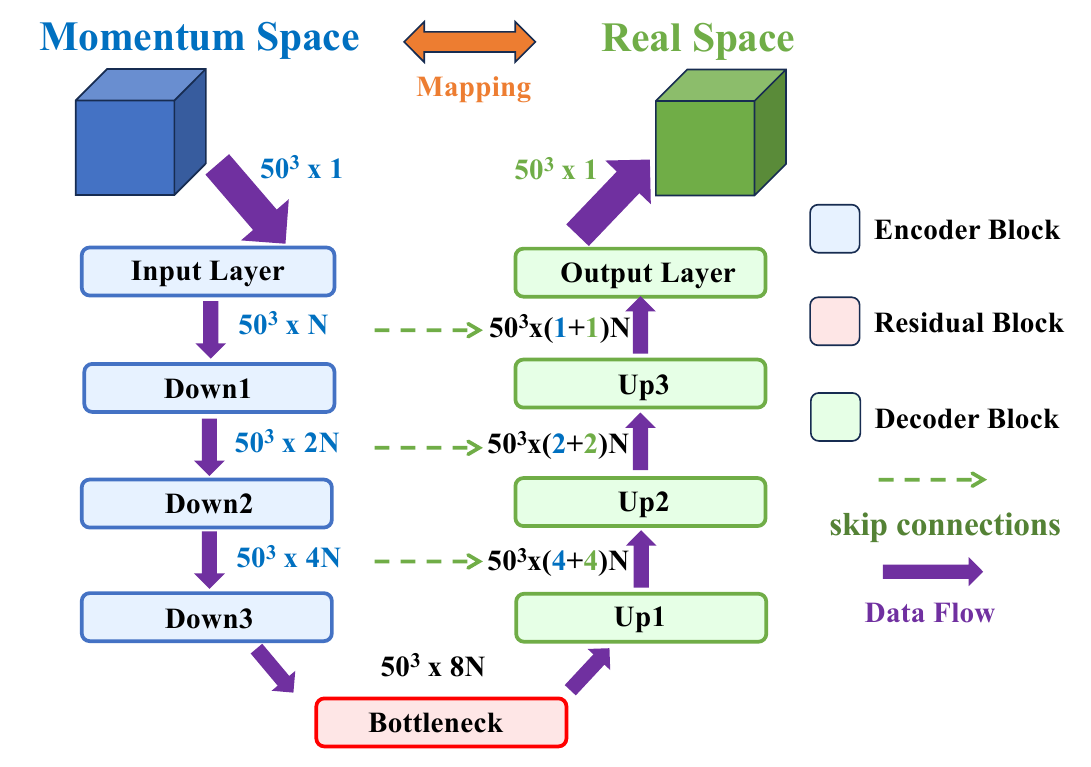}
	\caption{Schematic diagram of the deep neural network.}
	\label{Fig1}
\end{figure}

In this work, we present a Deep Neural Network (DNN) framework that enables direct position-space reconstruction of polyatomic molecular geometries from experimental CEI data. Our methodology integrates Coulomb explosion molecular dynamics (CEMD) simulations with a neural network architecture to address the fundamental challenges in molecular structure inversion.
We adopt an attention-augmented 3D ResUNet backbone (3D encoder–decoder with skip connections, residual blocks, GroupNorm) for volumetric-to-volumetric regression. 
The schematic diagram of the DNN framework is shown in Fig. \ref{Fig1}. The physics-informed neural network is trained to learn the mapping between momentum-space and real-space using corresponding images. The loss function of the DNN framework consists of two core components: Kullback–Leibler (KL) divergence and a boundary penalty. The former enforces global similarity between the predicted and target histograms by directly comparing their normalized probability distributions-encouraging the model to reproduce key features critical for molecular structure retrieval from Coulomb explosion imaging data. The latter acts as a physics-informed regularization term that mitigates unphysical boundary artifacts and promotes physically plausible solutions. By suppressing activations near the grid edges, it ensures the predicted distribution remains well-contained within the defined physical domain, avoiding spurious structures that violate the spatial constraints of real molecular geometries. The efficacy of the framework is demonstrated through successful reconstruction of CHF$_3$ molecular geometry from strong-field-induced CEI experimental data. 
This breakthrough establishes a foundation for gas-phase molecular structure determination and its inherent compatibility with time-resolved measurements opens new avenues for tracking ultrafast molecular dynamics.

The database for training the DNN framework was generated through CEMD simulations that numerically solve the classical equations of motion driven by the Coulombic field. These simulations assume instantaneous molecular ionization under the ultrafast laser field. To account for quantum zero-point fluctuations, the initial atomic positions and velocities were sampled from a Wigner phase-space distribution of the molecular ground state which was optimized by the $\omega$B97XD/aug-cc-pvtz method \cite{g16}. Furthermore, molecular geometries were systematically scaled to comprehensively cover the experimentally accessible configurations. This scaling protocol enables the momentum-space features observed in experiment to be definitively correlated with theoretical predications. 
Totally, 10 millions of trajectories are performed through CEMD simulations to generate the training database.
Ultimately, a robust machine learning-based inversion framework was established for reconstructing  three-dimensional molecular geometries from momentum-space data. 

Our DNN framework can operate in two distinct modes. The first mode, absolute geometry reconstruction, is designed to retrieve absolute three-dimensional structure, encompassing both steric configurations and bond lengths. This mode relies on training databases that explicitly encode the relationship between the absolute geometry and the corresponding momentum-space image, which demands substantial computational resources as the molecule’s size increases. The second mode maps the normalized geometry of a molecule to the normalized momentum-space image, where only the steric configuration is retrieved. In this scenario, the required computational resources remain comparable across all molecular sizes. The bond lengths, on the other hand, can be determined by comparing experimental and simulated kinetic energy releases (KERs), which are inversely proportional to the molecular bond lengths. In this work, the normalized mode is performed and we focus on reconstruction of the steric configurations. The grid precision for the neural network training is 50$\times$50$\times$50 (50$^3$) for every layer.

\begin{figure}[!t]
	\centering
	\includegraphics[width=7.5cm]{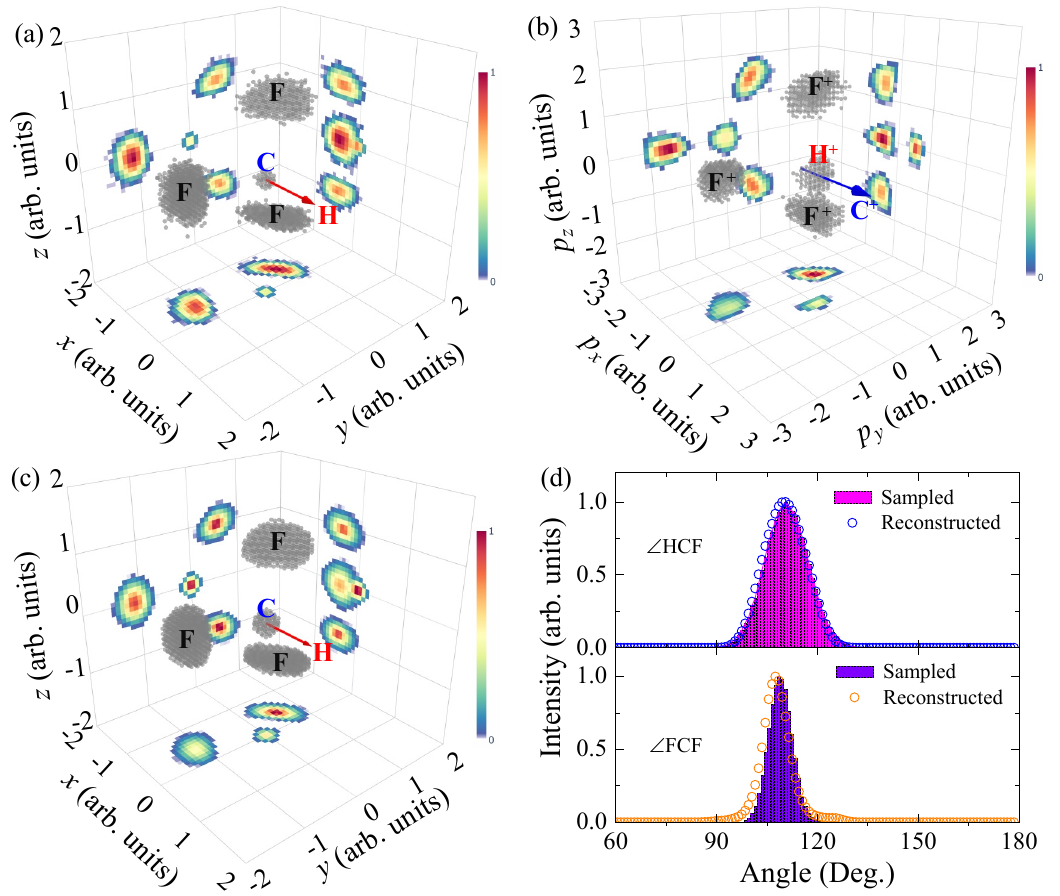}
	\caption{(a) The sampled geometries of CHF$_3$ from Wigner distribution. (b) Three-dimsnsional momentum-space Newton diagram by the CEI simulation. (c) The reconstructed position-space image of CHF$_3$ from the simulated momentum-space Newton diagram. (d) Comparsion bewteen the angular distribution of the sampled and reconstructed $\measuredangle$HCF and $\measuredangle$FCF angles of CHF$_3$. All of the Newton diagrams are presented in normalized scale and the angular distributions are normalized by their peaks.}
	\label{Fig2}
\end{figure}

The sampled position-space geometries from Wigner distribution of CHF$_3$ ground state are illustrated in Fig. \ref{Fig2} (a) where the bond lengthes are normalized to C-H bond. Such position-space Newton diagram is constructed by first rotating the vector of hydrogen to align with the $x$-axis and then rotating the vecotor sum of two C-F bonds onto the positive half of $xy$ plane. 
Within the Newton diagram, the three-dimensional (3D) molecular configuration is represented by the gray scatter points. The corresponding two-dimensional projections onto three orthogonal planes are illustrated by the colorful plots. For the simulated normalized momentum-space Newton diagram of CHF$_3$, as shown in Fig. \ref{Fig2} (b), the momenta of C$^+$ are rotated to the $x$-axis. Subsequently, the momentum sum of the two F$^+$ ions is oriented along the positive half of the $p_{xy}$ plane. All momenta are normalized to that of C$^+$. The proton carries less momenta during the Coulomb explosion process, thereby positioning itself near the center of the momentum-space Newton diagram. 
The reconstructed position-space distribution is shown in Fig. \ref{Fig2} (c). When comparing the positions and relative orientations of all atoms, the reconstructed steric conformation aligns closely with the sampled geometry distribution, as evident from the comparison between Fig. \ref{Fig2} (a) and (c). For a quantative validation, we analyze the angular distributions of angles $\measuredangle$HCF and $\measuredangle$FCF. As shown in Fig. \ref{Fig2} (d), the profiles of the reconstructed angular distributions agree well with the sampled ones. The obtained peak values are 110.5$\degree$ and 107.5$\degree$ for $\measuredangle$HCF and $\measuredangle$FCF, respectively, in line with the ground state of CHF$_3$, $\measuredangle$HCF  =  110.4 $\degree$, and $\measuredangle$FCF = 108.5 $\degree$.

To provide a global, geometry-level scalar assessment that complements the angle distributions, we quantify the proximity of CHF$_3$ geometries to the ideal $C_{3v}$ symmetry by the continuous symmetry measure (CSM) $S_{C_{3v}}$ \cite{Zabrodsky1993CSM2}. For a set of atomic coordinates $\{\mathbf{r}_i\}$, the CSM with respect to a symmetry group $G$ is defined as
\begin{equation}
 S_{G}=\min_{\mathcal{R},\,\{\mathbf{r}^{\,G}_i\}}\frac{\sum_i \left\lVert \mathbf{r}_i-\mathcal{R}\,\mathbf{r}^{\,G}_i \right\rVert^2}{\sum_i \left\lVert \mathbf{r}_i \right\rVert^2},
\end{equation}
where $\{\mathbf{r}^{\,G}_i\}$ is the closest $G$-symmetric configuration to $\{\mathbf{r}_i\}$, and $\mathcal{R}$ accounts for optimal rigid rotations/translations. $S_G\in[0,1]$ with smaller value represents higher symmetry (zero for an ideal $G$-symmetric structure).
For the sampled and reconstructed position-space ensembles, the obtained $S_{C_{3v}}$ = 0.003 and  0.004, respectively, indicating that the reconstruction preserves overall $C_{3v}$ symmetry. This scalar metric corroborates the agreement observed in the angular distributions which reflects the molecular symmetry. 

\begin{figure}[htbp]
	\centering
	\includegraphics[width=7.5cm]{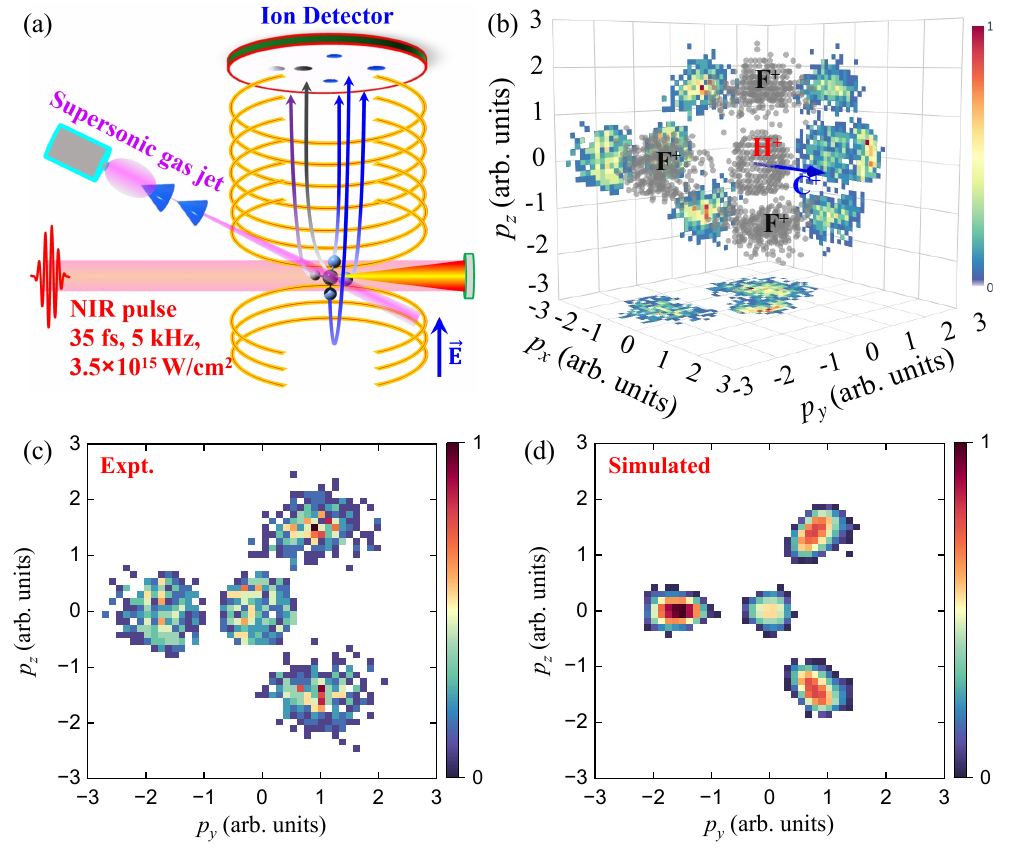}
	\caption{(a) Schematic of the experimental setup. (b) Experimental momentum-space Newton plot of CHF$_3$ measuring all 5 ions in coincidence. The the gray scatter points represent the 3D molecular configuration and the corresponding 2D projections onto three planes are illustrated by the colorful plots.
	 Comparing the experimental (c) and simulated (d) Newton plots on the $p_{yz}$ plane.} 
	\label{Fig3}
\end{figure}

An ultrafast strong-field driven CEI experiment was conducted utilizing a COLTRIMS reaction microscope \cite{Dorner2000PhysRep, Ullrich_2003} in combination with a Ti:Sapphire chirped-pulse amplification laser system, as schematically shown in Fig. \ref{Fig3} (a). The wavelength, repetition frequency, and width of the laser pulse were $\lambda$ = 800 nm, 5 kHz, and 35 fs,  respectively. The laser beam was backward-focused onto a molecular beam using a spherical concave mirror (focal length: 75 mm), achieving a peak intensity of approximately $3.5 \times 10^{15}\text{W/cm}^{2}$.  
Under such strong-field ionization conditions, the target molecule is rapidly stripped of multiple electrons, producing a highly charged ion that subsequently undergoes Coulomb explosion.
A supersonic jet expansion system delivered the CHF$_3$ molecular beam into the ultrahigh vacuum chamber through a 20 $\micro$m diameter nozzle, with subsequent collimation by two skimmers. Ionized fragments were accelerated by a uniform electric field (100 V/cm) and projected onto a multi-hit time- and position-sensitive delay-line detector (active area: 100 mm). The three-dimensional momentum vectors of all charged fragments were reconstructed from detected spatial coordinates ($x,~y$) and TOF ($t$) data. The experimental setup maintained a count rate of ~2 kHz, deliberately lower than the laser repetition rate to ensure high signal-to-noise ratios in momentum-resolved measurements.

The experimental momentum space Newton plot of CHF$_3$ derived from measuring the five-ion coincidence, i.e. H$^+$+C$^+$+3F$^+$, provides a critical validation of our DNN framework's ability to decode complex CEI data into position-space molecular geometries. This five-ion channel is particularly significant because inclusion of all five fragment ions ensures that the repulsive potentials are purely Coulombic during the explosion, and the reconstructed geometry retains maximal spatial information, effectively avoiding ambiguities that often plague CEI due to partial fragmentation or missing ion signals.
The experimental momentum-space Newton plot of the five ions in Fig. \ref{Fig3} (b) reveals distinct features that directly correlate with the molecular structure of CHF$_3$. As can be seen in the figure, the C$^+$ ion is positioned along the $x$-axis, while the three fluorine ions (F$^+$) exhibit a tetrahedral orientation opposite to C$^+$ ion. This symmetric pattern is a hallmark of the tetrahedral geometry of CHF$_3$. The proton appears at the center of the plot due to its smaller momenta compared to the heavier atoms.
The observed experimental momentum-space Newton plot exhibits consistency with the simulated counterpart, as demonstrated by the direct comparison in Fig. \ref{Fig2} (b) and \ref{Fig3} (b). 
To further validate this alignment and have finer structural insights, we perform two-dimensional projections of the momentum-space Newton plot onto the $p_{yz}$ plane, as shown in Fig. \ref{Fig3} (c) and (d) for experimental and simulated data, respectively. Both the patterns (angles between momentum vectors of the ions) and the magnitude of momentum vectors for all detected ions show excellent agreement between the experimental measurements and simulations.

\begin{figure}[htbp]
	\centering
	\includegraphics[width=7.5cm]{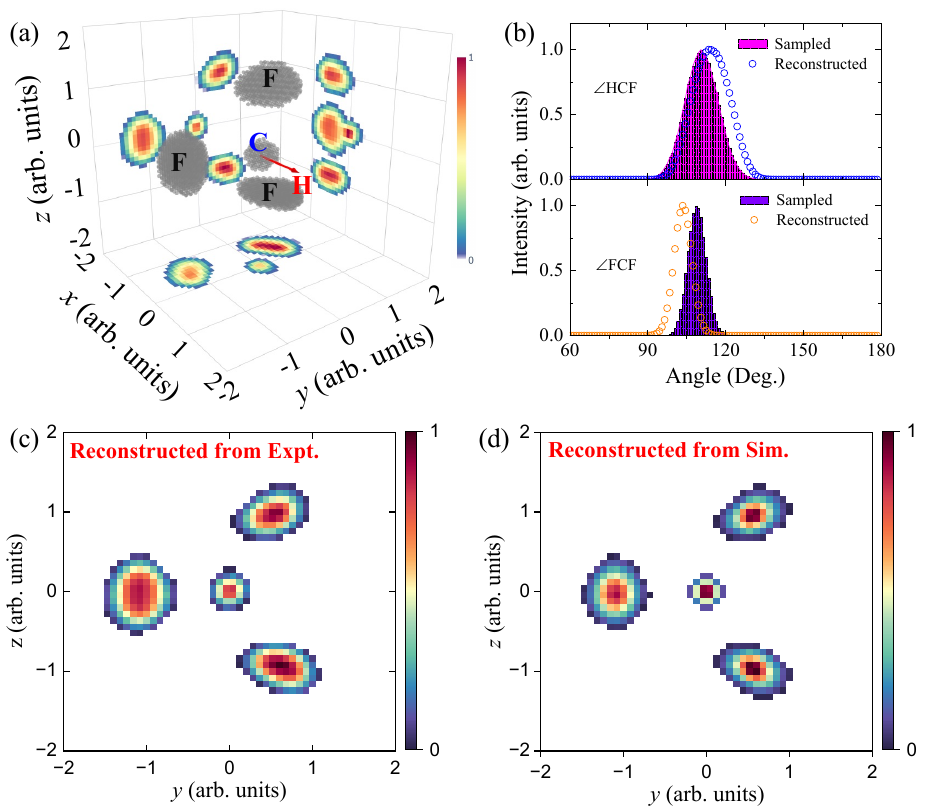}
	\caption{(a) The position-space goemetry distribution of CHF$_3$ retrieved from the momentum-space normalized Newton plot. (b) $\measuredangle$HCF and $\measuredangle$FCF angular distributions. Restructed (c) and sampled (d) position distributions of three fluorine atoms.} 
	\label{Fig4}
\end{figure}

The reconstructed position-space molecular geometry of CHF$_3$ is visualized in Fig. \ref{Fig4} (a), exhibiting strong consistency with the reference structure depicted in Fig. \ref{Fig2} (a), confirming the validity of our DNN-based reconstruction in capturing the three-dimensional configuration.  This agreement is further corroborated by the two-dimensional projections onto the $yz$-plane, presented in Fig. \ref{Fig4} (c) and (d) for the experimental and simulated geometries, respectively. Notably, the spatial distribution of fragment ions in both representations shows remarkable alignment in terms of angles and magnitudes, underscoring the reliability of the momentum-to-position mapping learned by the neural network. The experimental CSM is $S_{C_{3v}}$ = 0.006 refecting a very good  $C_{3v}$ symmetry. Fig. \ref{Fig4} (b) displays the experimental angular distributions with reconstructed peaks at $\measuredangle$HCF  =  103.5$\degree$, and $\measuredangle$FCF = 113.5$\degree$, where $\measuredangle$HCF exceeds and $\measuredangle$FCF falls below the equilibrium angles of equilibrium CHF$_3$ due to laser-induced geometry relaxation from sequential ionization of five electrons. This process proceeds along the exact molecular potential energy surface, leading to deviation from purely Coulombic repulsive behavior.

Regarding the bond lengths, the equalibrium C-H and C-F bond lengths in the CEMD simulation are 1.09~\AA~and~1.33~\AA~for C-H and C-F bonds, respectively. The experimental KER peak in this work is about 55 eV, which is lower than the result from the CEMD simulation (KER = 89.8 eV). This is because each fragment is assigned an integer charge in the CEMD simulation under the assumption of instantaneous ionization. The pure Coulomb potential deviates from the accurate molecular potential energy surface (PES). Consequently, the bond lengths reconstructed from the experimental data will be longer than their equilibrium counterparts.
Crucially, when normalizing the C–H internuclear distance to unity, the relative C–F bond length in the reconstructed geometry measures 1.35, aligning closely with the equilibrium bond length ratio (1.22).

To enhance the simulation accuracy, the charge-buildup duration induced by the strong-field laser pulse is incorporated \cite{Li_PRR2022}. Here, the molecular charge-buildup rate is characterized by a Gaussian profile with a width matching the laser pulse duration (FWHM = 35 fs). The total charge is uniformly allocated to each atom of the molecule. In this case, the simulated KER is approximately 70 eV, and the experimental KER is about 21\% lower than the simulated value.
By using the reconstructed bond angles and bond length ratio, and incorporating the charge-buildup CEMD simulation, we determined the experimental bond lengths to be C-H = 1.31~\AA~and C-F = 1.75~\AA, which are about 25\% longer on average than the equilibrium bond lengths.
Recently, Li et al. \cite{XLi2025Arxiv} proposed a diffusion-based transformer neural network that can reconstruct unknown molecular geometries from ion-momentum distributions of Coulomb explosion imaging, with a mean absolute error below one Bohr radius (half the length of a typical chemical bond). Our normalized model of the DNN framework offers a more computationally efficient solution.

In summary, the synergistic integration of strong-laser-field CEI experiments and a physics-informed training database for the deep neural network framework enables high-fidelity three-dimensional structural reconstruction of gas-phase molecules. In experimental, the Coulomb explosion of CHF$_3$ is triggered by intense femtosecond-laser pulses and captured via a COLTRIMS reaction microscope. Through multibody coincidence measurements, the 3D momentum-space image of the molecule is constructed using the Newton plot method. Complementarily, the simulation-derived training database for the DNN framework is generated by numerically solving classical equations of motion governed by Coulombic forces. This process yields not only momentum-space 3D images but also position-space structural representations of CHF$_3$, forming a dual-domain dataset that anchors the training of neural network. The steric configuration reconstructed from experimental 3D Newton plots demonstrates agreement with the sampled geometries from the Wigner distributions.

A common challenge in strong-laser-field experiments is the charge-buildup duration and molecular structural distortion, both induced by the finite laser pulse duration. This leads to a lower experimental kinetic energy release (KER) compared to ideal Coulomb explosion scenarios, with the resulting predicted molecular bond lengths being longer than equilibrium values. To mitigate this issue, ultrashort laser pulses—for example, through subsequent temporal compression or the utilization of ultrashort X-ray free-electron lasers (XFELs)—are required to minimize molecular structural perturbations during ionization.
Notably, the flexibility of the DNN framework allows for substitution of the Coulomb explosion simulation mode with $ab~initio$ molecular dynamics (AIMD) simulations, enabling direct incorporation of quantum mechanical interactions. This adaptability positions the framework as a versatile tool for future applications in both classical CEI and quantum-based structure imaging of molecules. The  transformer neural network by Li $et~al$ \cite{XLi2025Arxiv} was trained by 16 and 4 A100 GPUs for each of the training steps. The deep neural network presented here achieves high computational efficiency while running on only two NVIDIA RTX4090 GPUs. This approach can be extended to large molecular systems and has potential in time-resolved molecular imaging, bridging experimental observations with physics-based simulations to achieve reliable structural insights at femtosecond resolution.

This work was jointly supported by the National Natural Science Foundation of China (Grant No. 12450404, 12127804), the National Key Research and Development Program of China (Grant No. 2022YFA1602502), the CAS Pioneer Hundred Talents Program (No. KJ2030007007), and the USTC Research Funds of the Double First-Class Initiative (No. YD2030002020). The CEMD calculations in this paper were performed on the supercomputing system in the Supercomputing Center of University of Science and Technology of China.

\bibliography{ref}

\end{document}